\documentclass[12pt,a4paper]{article}

\usepackage[T1]{fontenc}
\usepackage{graphicx}
\usepackage{epsfig}
\usepackage{amsmath}
\usepackage{amssymb}
\usepackage{latexsym}

\usepackage[usenames]{color}

\begin{document}
\textwidth=135mm
 \textheight=200mm
\begin{center}
{\bfseries Hydrokinetic predictions for femtoscopy scales in A+A collisions in the light of recent ALICE LHC results
\footnote{{\small The entry is based on the talks given at the Sixth Workshop on Particle Correlations and Femtoscopy, BITP, Kiev, September 14 - 18,
2010 and GSI/EMMI Seminar, January 14, 2011}}}
\vskip 5mm
Yu.M. Sinyukov$^{\dag, \ddag}$ and Iu.A. Karpenko$^{\dag}$ 
\vskip 5mm
{\small {$^{\dag}$ \it Bogolyubov Institute for Theoretical Physics, Kiev, 03680, Ukraine}} \\
{\small {$^{\ddag}$ \it ExtreMe Matter Institute EMMI, GSI
Helmholtz Zentrum f\"ur
  Schwerionenforschung,
D-64291 Darmstadt, Germany}} \\
\end{center}
\vskip 5mm
\centerline{\bf Abstract}
A study of energy behavior of the pion spectra and interferometry
scales is carried out for the top SPS, RHIC and for LHC energies within
the hydrokinetic approach. The main mechanisms that lead to the paradoxical,
at first sight, dependence of the interferometry scales with an energy growth, in particular, a decrease $R_{out}/R_{side}$ ratio, are exposed.
The hydrokinetic predictions for the HBT radii at  LHC energies are compared with the recent results of the ALICE experiment.
\vskip 10mm
\section{Introduction}

In anticipation of start of the LHC experiments there were presented many different theoretical views on the underlying physics at so large collision energies as well as  numerous predictions for observables - a big collection of the predictions for heavy ion collision was assembled in \cite{LHCpred}. The similar took place on the eve of the RHIC experiments. At that ``pre-RHIC'' time the future results on the correlation femtoscopy of particles, that is the topic of this entry, were expecting with great interest. One of the reason was a hope to find the interferometry signature of the quark-gluon plasma (QGP). A very large value of the 
ratio of the two transverse interferometry radii, $R_{out}$ to $R_{side}$, was predicted as a signal of the QGP formation \cite{Bertsch}. While the $R_{side}$ radius
is associated with the transverse homogeneity length \cite{Sin}, the $R_{out}$
includes besides that also additional contributions, in particular,
the one which is related to a duration of the pion emission. Since
the lifetime of the systems obviously should grow with collision
energy, if it is accompanied by an increase of the initial energy
density and/or by a softening of the equation of state due to phase
transition between hadron matter and quark-gluon plasma (QGP), 
the duration of pion emission should also grow with energy and so
$R_{out}/R_{side}$ ratio could increase.

The RHIC experiments brought an unexpected result: the ratio \\
$R_{out}/R_{side}\approx 1$ is similar or even smaller than at SPS.
The another surprise was the absolute values of the radii.  Naively it was expected that when the energy of colliding nuclei
increases, the pion interferometry volume $V_{int}$ - product of
the interferometry radii in three orthogonal directions - will rise
at the same maximal centrality for Pb+Pb and Au+Au collisions just
proportionally to $\frac{dN_{\pi}}{dy}$. However, when experiments
at RHIC starts, an increase of the interferometry volume with energy
turn out to be essentially smaller then if the  proportionality law takes place.
Both these unexpected results were called the RHIC HBT puzzle
\cite{HBTpuzzle}. 

During a long period this puzzle was not solved in hydrodynamic/hybrid models of A+A collisions which reproduce good the single particle transverse spectra and its axial anisotropy in non-central collisions described by the $v_2$ coefficients. Only a few years ago the main factors which allow one
to describe simultaneously the spectra and femtoscopic
scales at RHIC become clear. They are \cite{sin1}-\cite{Pratt}: a relatively hard equation of state because of crossover transition (instead of the 1st order one) between quark-gluon and hadron phases and due to
nonequilibrium composition of hadronic matter,
presence of prethermal transverse flows and their anisotropy developed to thermalization time,
an `additional portion' of the transverse flows
owing to the shear viscosity effect and fluctuation of initial
conditions. An account of these factors gives the possibility to describe good the pion and kaon spectra together with the femtoscopy data of RHIC within realistic freeze-out picture with a gradual decay of
nonequilibrium fluid into observed particles \cite{sin3}.

Now, when the heavy ion experiments at LHC already starts, and the ALICE Collaboration published the first results on the femtoscopy in A+A collisions at $\sqrt{s}=2.76$ TeV \cite{Alice}, the main question is whether an understanding of the physics responsible for the space-time matter evolution in Au+Au collisions at RHIC can be extrapolated to the LHC energies, or some new ``LHC HBT  puzzle'' is already apprehended just as it happened in the way from SPS to RHIC energies. In this note we describe the physical mechanisms responsible for the peculiarities of energy
dependence of the interferometry radii and therefore solving the RHIC HBT puzzle, present the quantitative predictions given for LHC  within hydrokinetic model earlier \cite{sin4}, compare them with the recent ALICE LHC results and make the corresponding inference.

\section{Hydro-kinetic approach to A+A collisions}

Let us briefly describe the main features of the HKM \cite{PRL,PRC}.
It incorporates hydrodynamical expansion of the systems formed in
\textit{A}+\textit{A} collisions and their dynamical decoupling
described by escape probabilities. 

{\it Initial conditions}--- Our results are all related to the
central rapidity slice where we use the boost-invariant Bjorken-like
initial condition. We consider the proper time of thermalization of
quark-gluon matter to be
$\tau_0=1$ fm/c, at present there is no theoretical arguments permitting smaller value.  The initial energy density in the transverse plane
is supposed to be Glauber-like \cite{Kolb}, i.e. is proportional to
the participant nucleon density  for Pb+Pb (SPS) and Au+Au (RHIC,
LHC) collisions with zero impact parameter. The height of the
distribution - the maximal initial energy density -
$\epsilon(r=0)=\epsilon_0$ is the fitting parameter. From analysis
of pion transverse spectra we choose it for the top SPS energy to be
$\epsilon_0 = 9$ GeV/fm$^3$ ($\langle\epsilon\rangle_0$ = 6.4
GeV/fm$^3$), for the top RHIC energy  $\epsilon_0 = 16.5$ GeV/fm$^3$
($\langle\epsilon\rangle_0$ = 11.6 GeV/fm$^3$). The brackets $<...>$
correspond to mean value over the distribution associated with the
Glauber transverse profile. We also demonstrate results at
$\epsilon_0 = 40$ GeV/fm$^3$ and $\epsilon_0 = 60$ GeV/fm$^3$. In hydrokinetic model $\epsilon_0 = 40$ GeV/fm$^3$ correspond to multiplicity of charged particles $dN_{ch}/d\eta \approx 1500$. We suppose that soon after thermalization the matter
created in A+A collision at energies considered is in the quark gluon
plasma (QGP) state. 

At the time of thermalization, $\tau_0=1$ fm/c, the system already
has developed collective transverse velocities \cite{sin1,JPG}. The
initial transverse rapidity profile is supposed to be linear in
radius $r_T$:
\begin{equation}
y_T=\alpha\frac{r_T}{R_T} \label{yT},
\end{equation}
where $\alpha$ is the second fitting parameter and
$R_T=\sqrt{<r_T^2>}$. Note that the fitting parameter $\alpha$
should absorbs also a positive correction for underestimated
resulting transverse flow since in this work we did not account in
direct way for the viscosity effects \cite{Teaney} neither at QGP
stage nor at hadronic one. In formalism of HKM \cite{PRC} the
viscosity effects at hadronic stage are incorporated in the
mechanisms of the back reaction of particle emission on hydrodynamic
evolution which we ignore in current calculations. Since the
corrections to transverse flows which depend on unknown viscosity
coefficients are unknown, we use fitting parameter $\alpha$ to
describe the "additional unknown portion" of flows, caused both
factors: by a developing of the pre-thermal flows and the viscosity
effects in quark-gluon plasma. The best fits of the pion transverse
spectra at SPS and RHIC are provided at $\alpha=0.194$ ($\langle v_T
\rangle= 0.178$) for SPS energies and $\alpha=0.28$ ($\langle
v_T\rangle=0.25$) for RHIC ones. The latter value we use also for
LHC energies aiming to analyze just influence of energy density
increase. 

{\it Equation of state}--- Following to Ref. \cite{PRC} we use at
high temperatures the EoS \cite{Laine} adjusted to the QCD lattice
data with the baryonic chemical potential $\mu_B =0$ and matched
with chemically equilibrated multi-component hadron resonance gas at
$T=175$ MeV. Such an EoS could be a good approximation for the RHIC
and LHC energies; as for the SPS energies we utilize it just  to
demonstrate the energy dependent mechanism of formation of the
space-time scales \footnote{a good description of the spectra and HBT radii at the SPS energies with realistic EoS within hydrokinetic model is presented in Ref.\cite{Toneev}}. We suppose the chemical freeze-out for the hadron
gas at $T_{ch}=165$ MeV \cite{PBM1}. It guarantees us the correct
particle number ratios for all quasi-stable particles (here we
calculate only pion observables) at least for RHIC. Below $T_{ch}$ a
composition of the hadron gas is changed only due to resonance
decays into expanding fluid. We include 359 hadron states made of u,
d, s quarks with masses up to 2.6 GeV. The EoS in this non
chemically equilibrated system depends now on particle number
densities $n_i$ of all the 359 particle species $i$:
$p=p(\epsilon,\{n_i\})$. Since the energy densities in expanding
system do not directly correlate with resonance decays, all the
variables in the EoS depend on space-time points and so an
evaluation of the EoS is incorporated in the hydrodynamic code. We
calculate the EoS below $T_{ch}$ in the Boltzmann approximation of
ideal multi-component hadron gas.

{\it Evolution}--- At the temperatures higher than $T_{ch}$ the
hydrodynamic evolution is related to the quark-gluon and hadron
phases which are in chemical equilibrium with zero baryonic chemical
potential. The evolution is described by the conservation law for
the energy-momentum tensor of perfect fluid:
\begin{equation}
\partial_\nu T^{\mu\nu}(x)=0
\label{conservation}
\end{equation}
At $T<T_{ch}$=165 MeV the system evolves as non chemically
equilibrated hadronic gas. The concept of the chemical freeze-out
implies that afterwards  only elastic collisions and resonance decays
take place because of relatively small
densities allied with a fast rate of expansion at the last stage.
Thus, in addition to (\ref{conservation}), the equations accounting
for the particle number conservation and resonance decays are added.
If one neglects the thermal motion of heavy resonances the equations
for particle densities $n_i(x)$ take the form:
\begin{equation}
    \partial_\mu(n_i(x) u^\mu(x))=-\Gamma_i n_i(x) + \sum\limits_j b_{ij}\Gamma_j
    n_j(x)
    \label{paricle_number_conservation}
\end{equation}
where $b_{ij}=B_{ij}N_{ij}$ denote the average number of i-th
particles coming from arbitrary decay of j-th resonance,
$B_{ij}=\Gamma_{ij}/\Gamma_{j,tot}$ is branching ratio, $N_{ij}$ is
a number of i-th particles produced in $j\rightarrow i$ decay
channel. We also can account for recombination in the processes of
resonance decays into expanding medium just by utilizing the
effective decay width $\Gamma_{i,eff}=\gamma\Gamma_i$.  We use
$\gamma = 0.75$ supposing thus that near 30\% of resonances are
recombining during the evolution. All the equations
(\ref{conservation}) and 359 equations
(\ref{paricle_number_conservation}) are solving simultaneously with
calculation of the  EoS, $p(x)=p(\epsilon(x),\{n_i(x)\})$, at each
point $x$.

{\it System's decoupling and spectra formation} --- During the
matter evolution, in fact, at $T\leq T_{ch}$, hadrons continuously
leave the system. Such a process is described by means of the
emission function $S(x,p)$ which is expressed for pions through the
{\it gain} term, $G_{\pi}(x,p)$, in Boltzmann equations and
the escape probabilities ${\cal P}_{\pi}(x,p)=\exp(-\int\limits_{t}^{\infty}%
dsR_{\pi+h}(s,{\bf r}+\frac{{\bf p}}{p^0}(s-t),p))$:
$S_{\pi}(x,p)=G_{\pi}(x,p){\cal P}_{\pi}(x,p)$ \cite{PRL,PRC}. For
pion emission in relaxation time approximation $G_{\pi}\approx
f_{\pi}R_{\pi+h}+G_{H\rightarrow\pi}$ where $f_{\pi}(x,p)$ is the
pion Bose-Einstein phase-space distribution, $R_{\pi+h}(x,p)$ is the
total collision rate of the pion, carrying momentum $p$, with all
the hadrons $h$ in the system in a vicinity of point $x$,  the term
$G_{H\rightarrow\pi}$ describes an inflow of the pions into
phase-space point $(x,p)$ due to the resonance decays. It is
calculated according to the kinematics of decays with simplification
that the spectral function of the resonance $H$ is
$\delta(p^2-\langle m_H\rangle^2)$. The cross-sections in the
hadronic gas, that determine via the collision rate $R_{\pi+h}$ the
escape probabilities ${\cal P}(x,p)$ and emission function $S(x,p)$,
are calculated in accordance with the UrQMD method \cite{UrQMD}. The
spectra and correlation functions are found from the emission
function $S$ in the standard way (see, e.g., \cite{PRL}).
\begin{figure*}[htb]
\centering
 \includegraphics[scale=0.25]{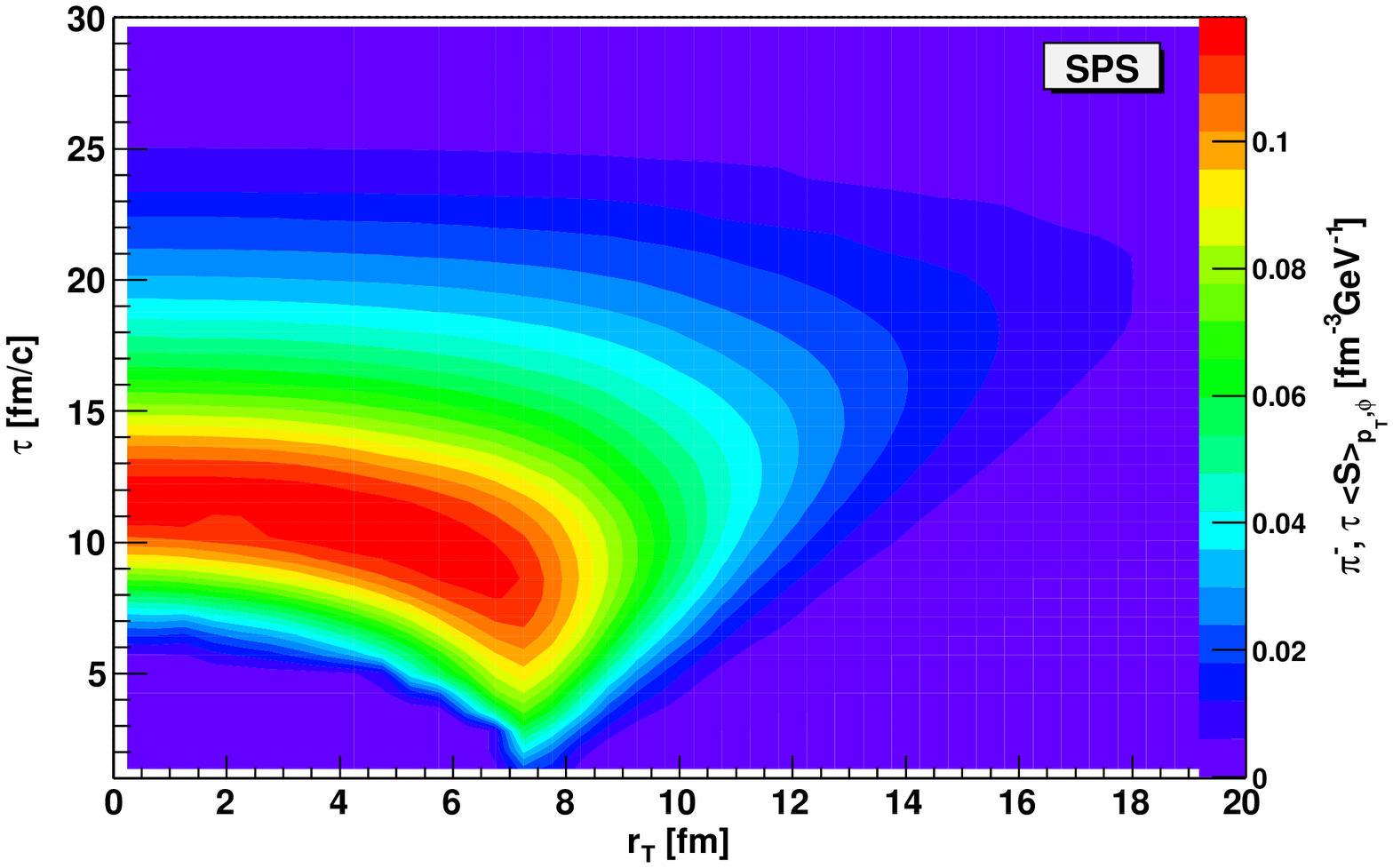}%
 \includegraphics[scale=0.25]{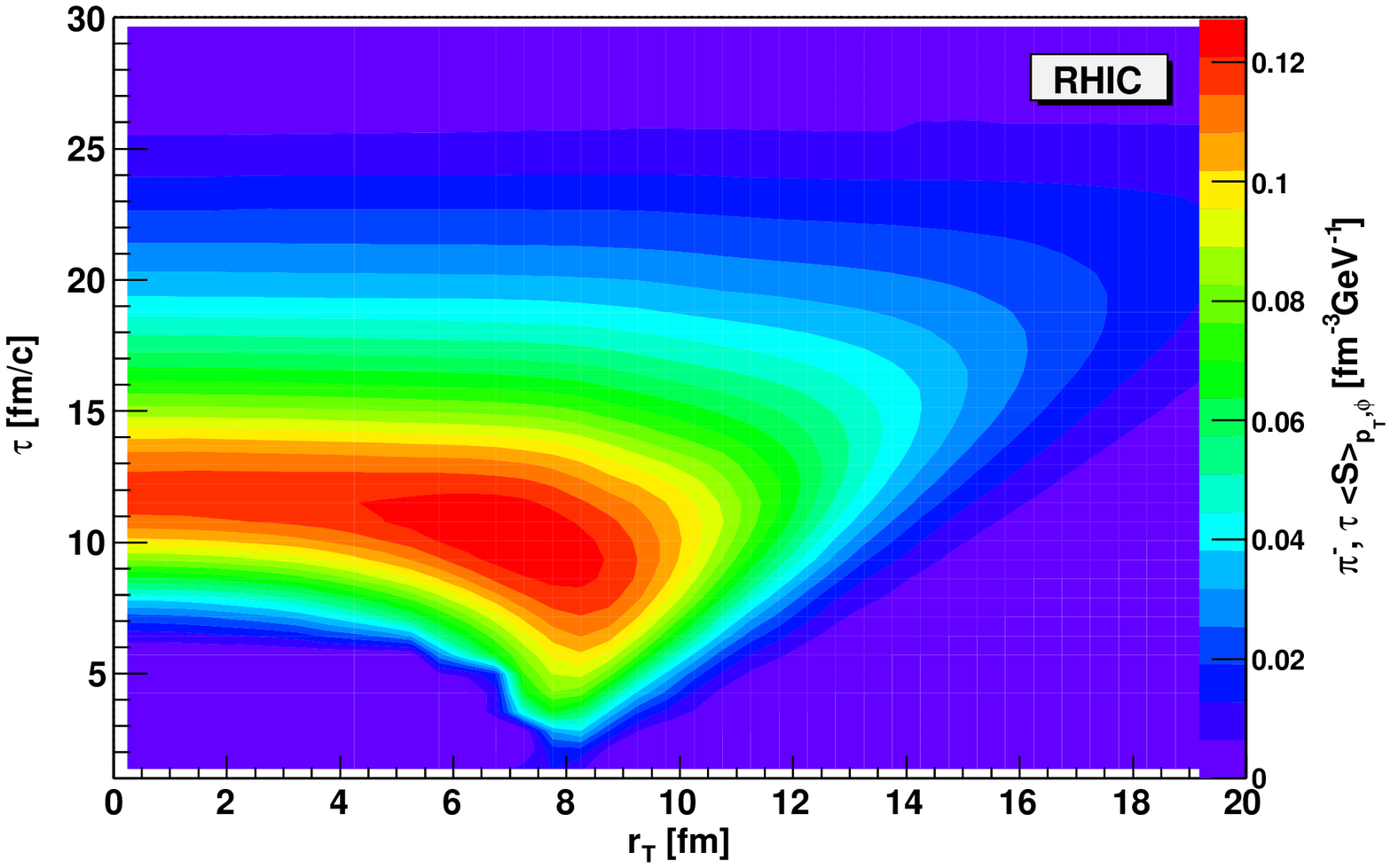}%
 \includegraphics[scale=0.25]{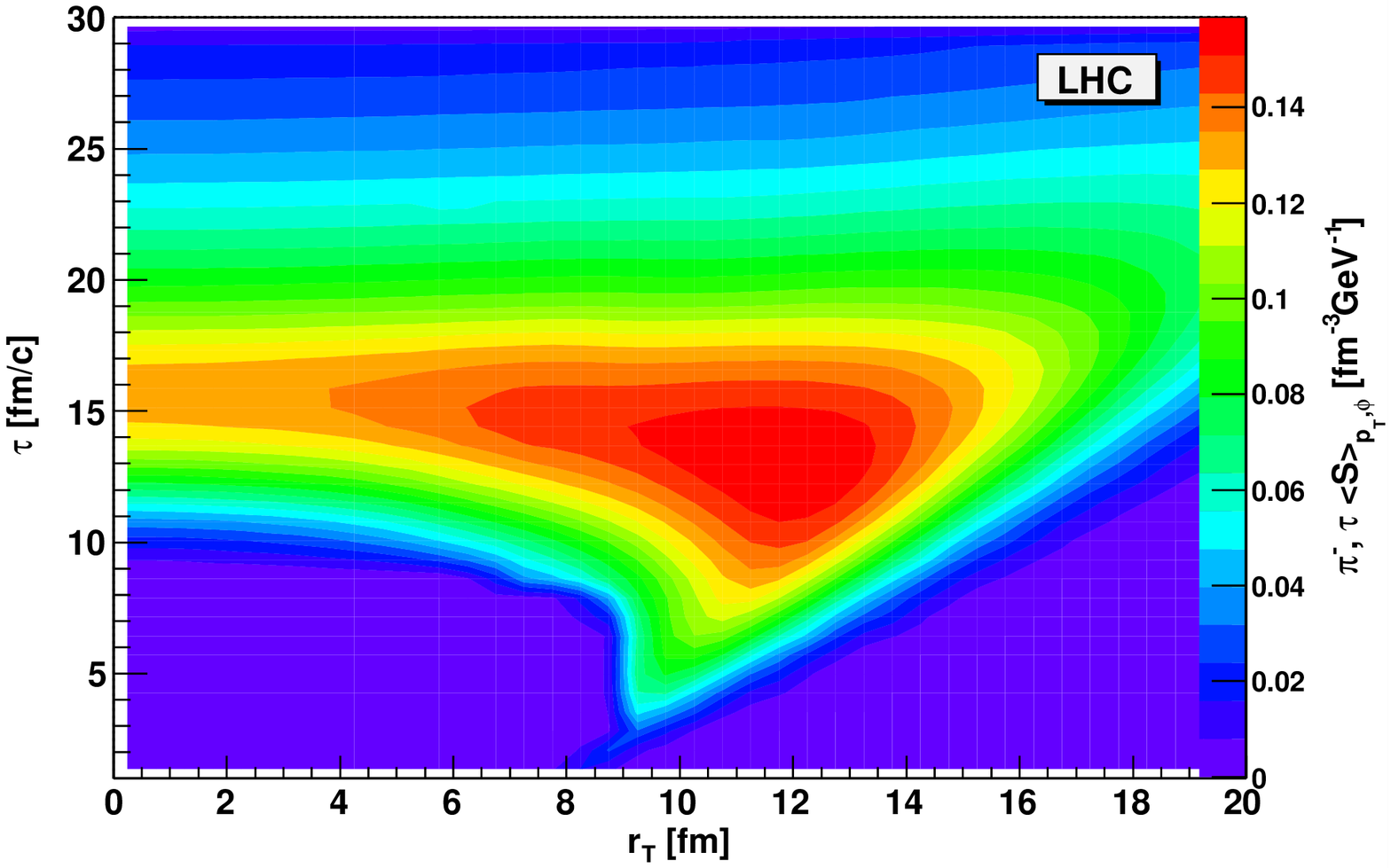}
 \includegraphics[scale=0.25]{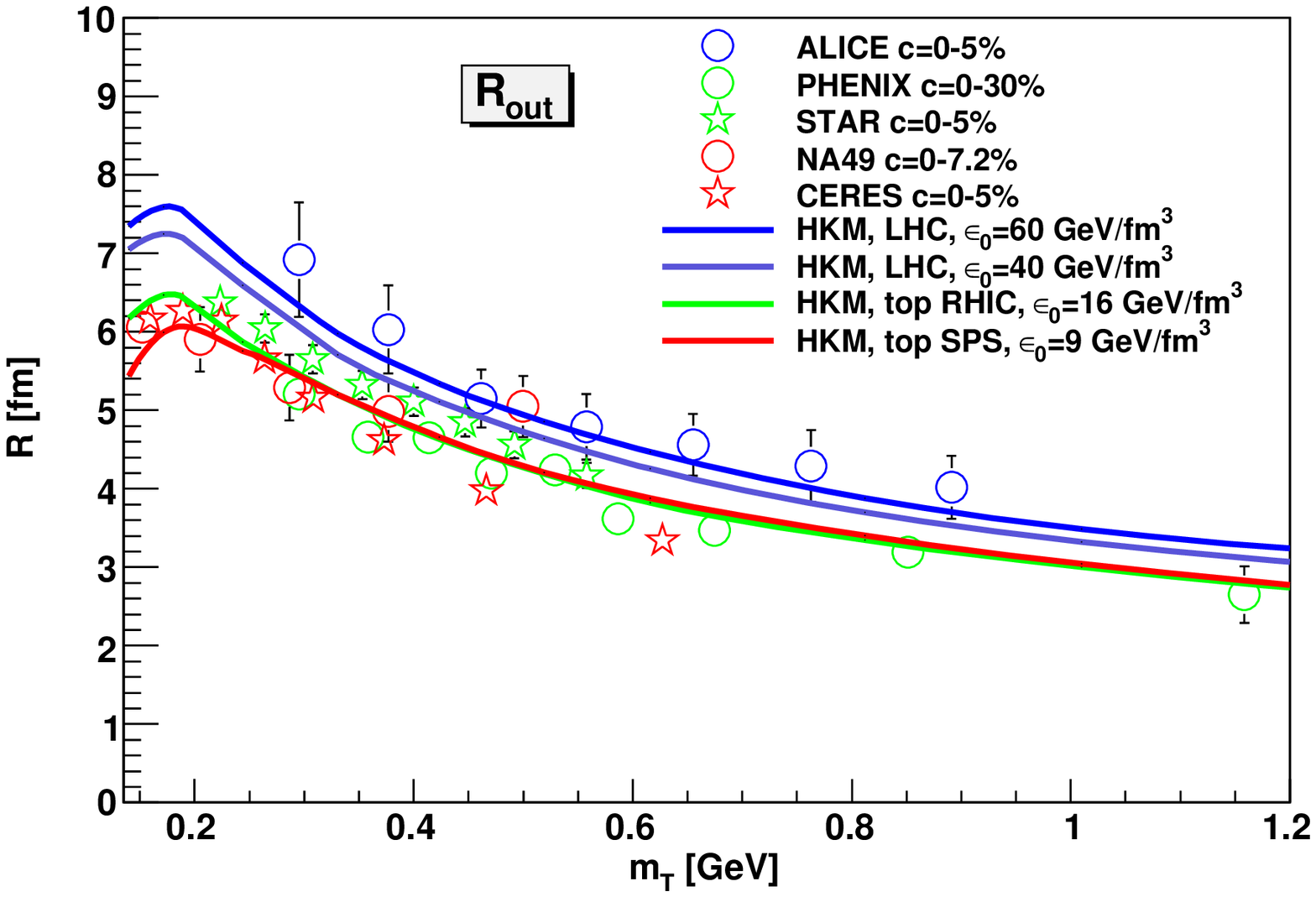}%
 \includegraphics[scale=0.25]{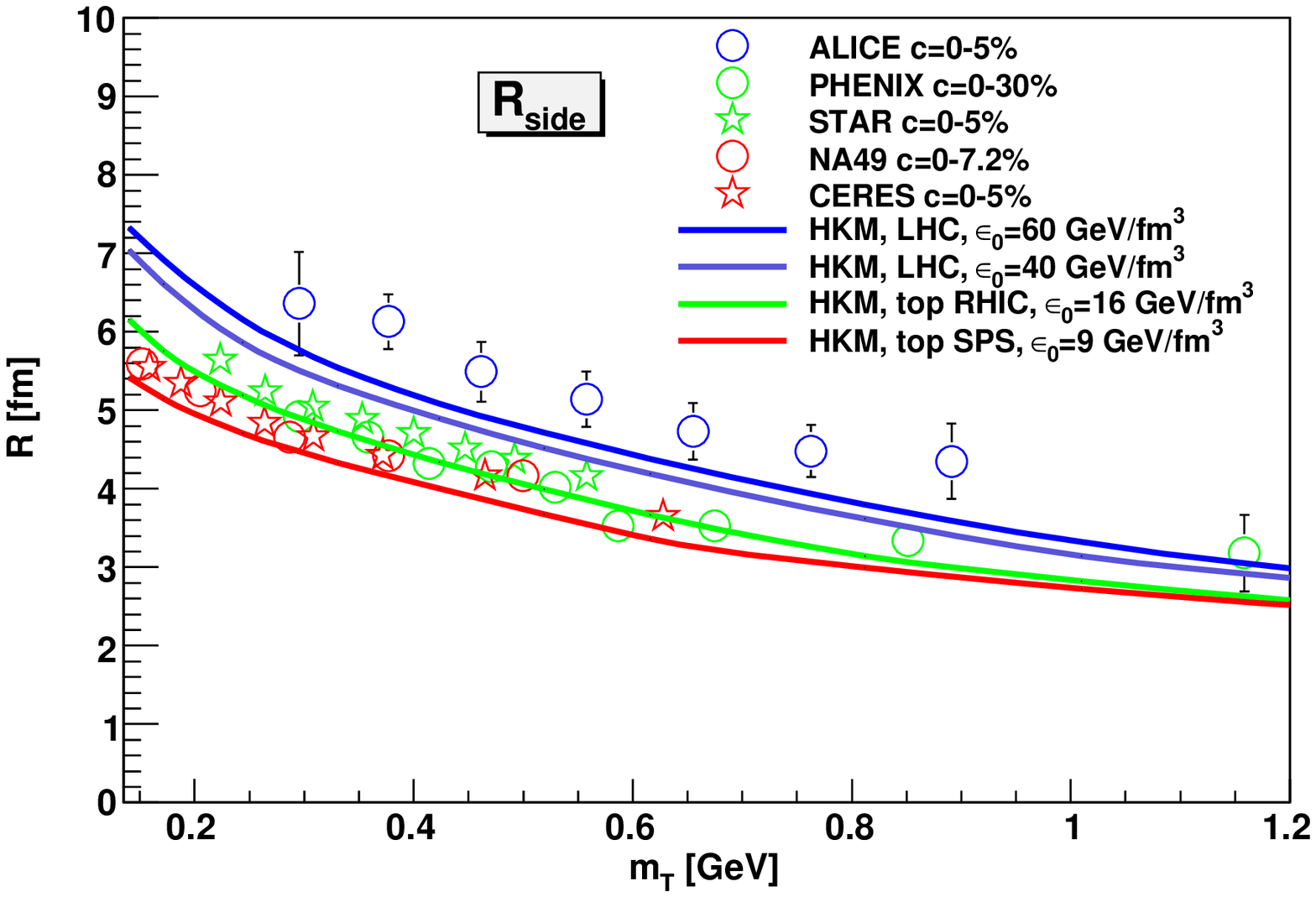}%
 \includegraphics[scale=0.25]{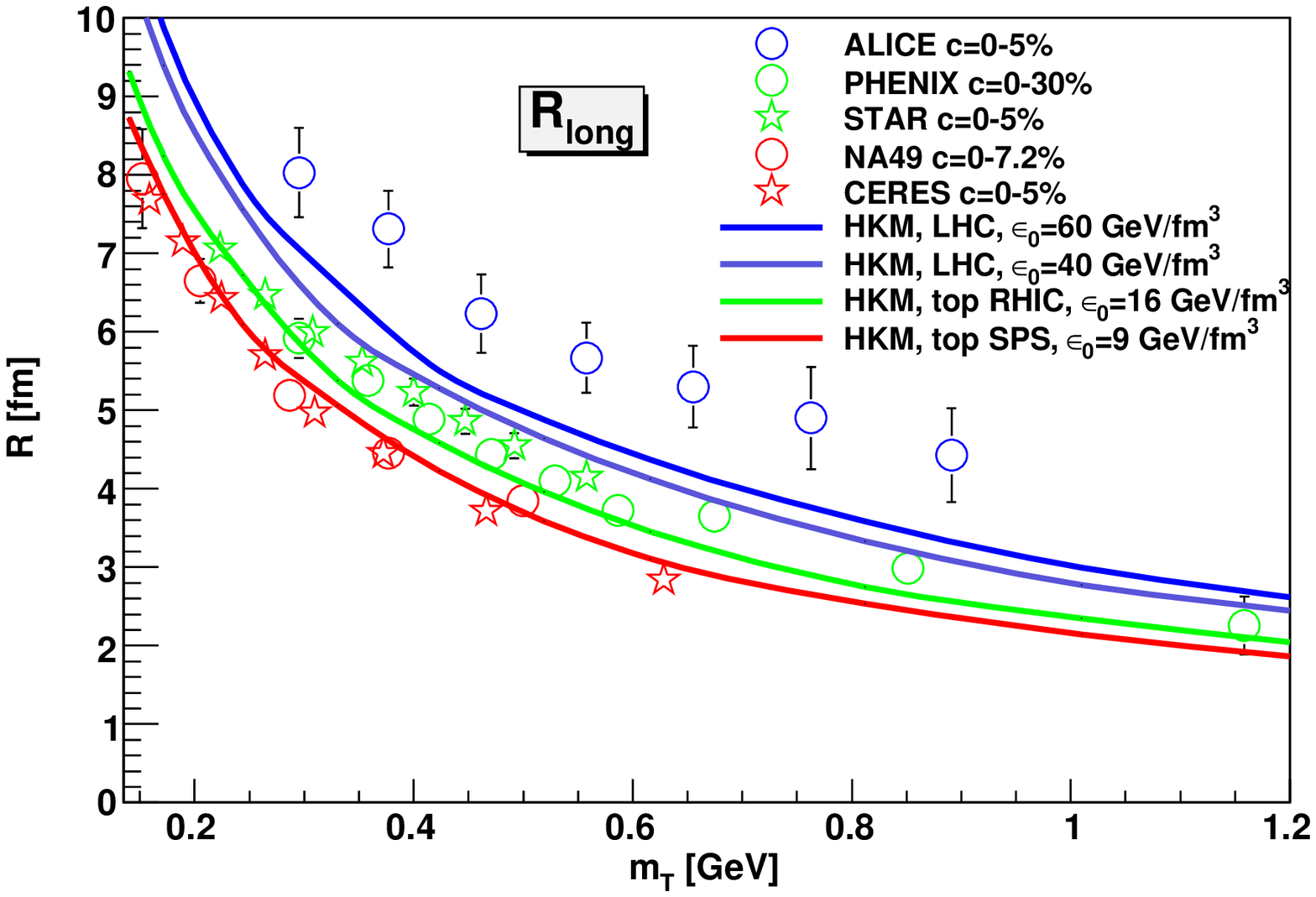}
 \includegraphics[scale=0.25]{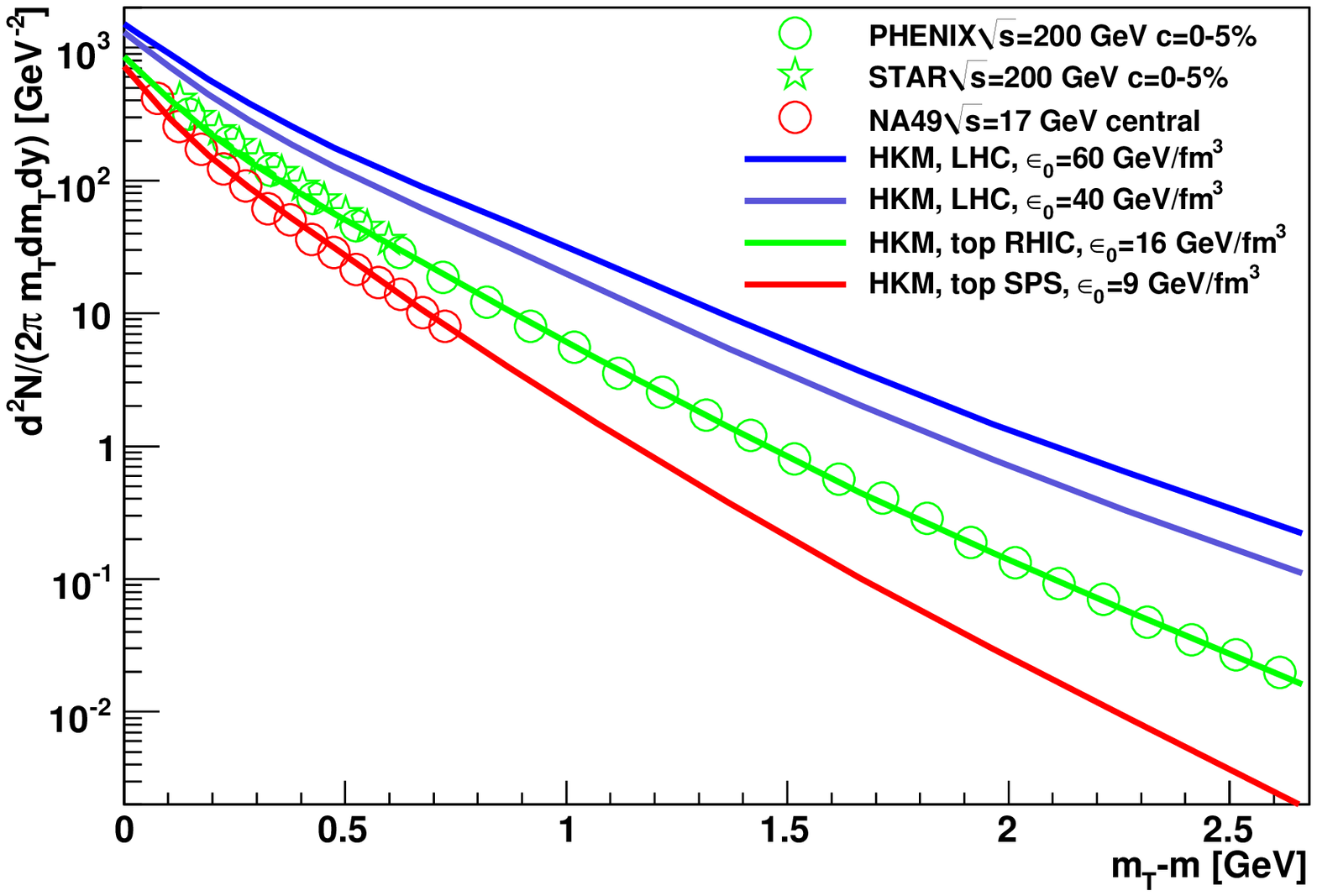}%
 \includegraphics[scale=0.25]{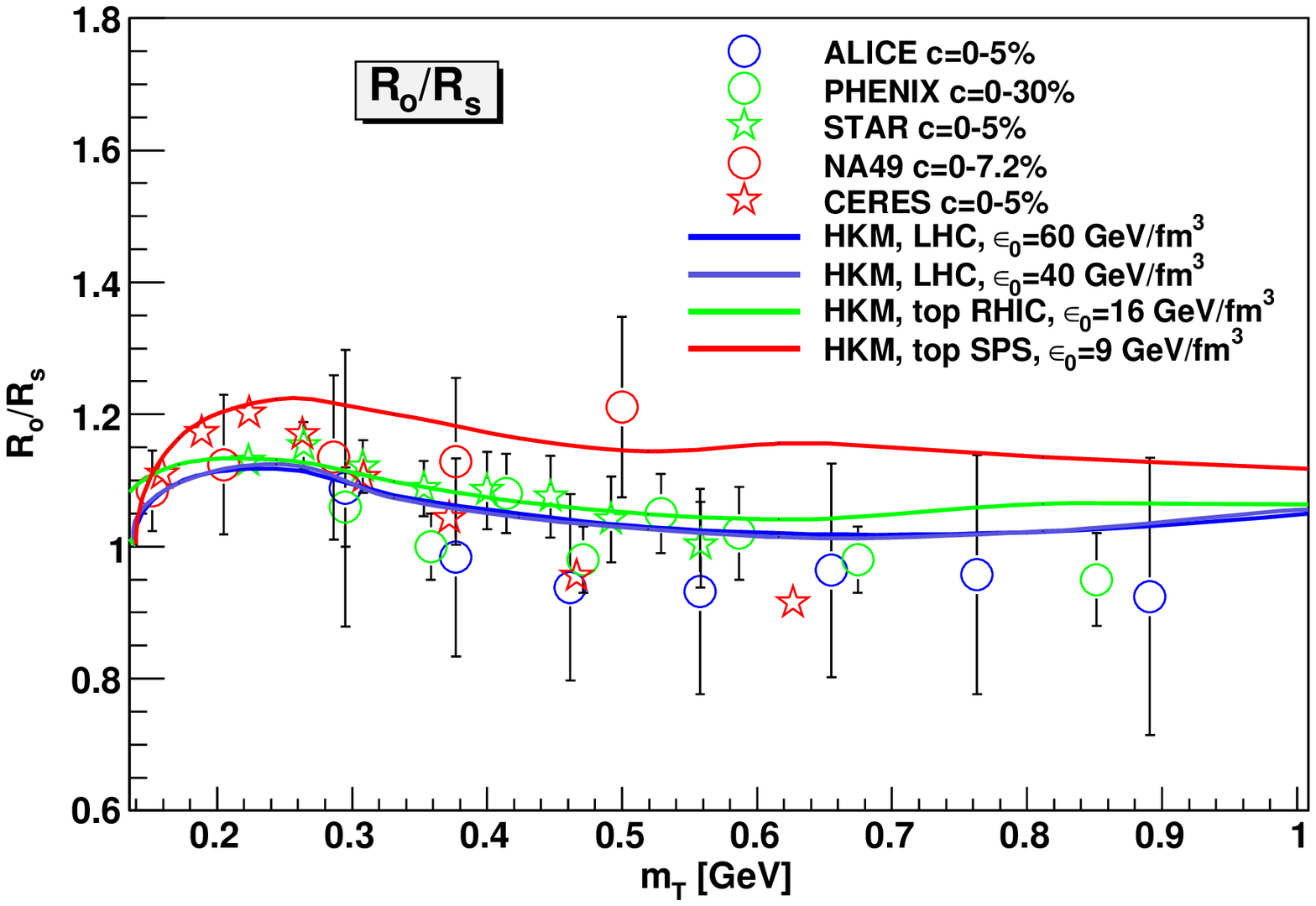}

 \vspace{-0.1in}
\caption{The $p_T$-integrated emission functions of negative pions
for the top SPS, RHIC and LHC energies (top); the interferometry
radii (middle) $R_{out}/R_{side}$ ratio and transverse momentum
spectra (bottom) of negative pions  at different energy densities,
all calculated in HKM model. The experimental data are taken from
CERES \cite{ceres} and NA-49 Collaborations \cite{na49-spectra, na49-hbt} (SPS CERN),
STAR \cite{star-spectra, star-hbt} and PHENIX \cite{phenix-spectra, phenix-hbt} Collaborations (RHIC BNL) and ALICE  Collaboration (LHC, CERN)\cite{Alice}.} 
\end{figure*}

\section{Results and conclusions}
The pion emission function per unit (central) rapidity, integrated
over azimuthal angular and transverse momenta, is presented in Fig.
1 for the top SPS, RHIC and LHC energies as a function of transverse
radius $r$ and proper time $\tau$.  The two fitting parameters
 and $\langle v_T \rangle$ are fixed as discussed above,
$\epsilon_0$ is also marked in figures. The pion transverse momentum spectrum, its
slope as well as the absolute value, and the interferometry radii,
including $R_{out}$ to $R_{side}$ ratio, are in a good agreement
with the experimental data both for the top SPS and RHIC energies.

As one can see particle emission lasts a total lifetime of the
fireballs; in the cental part, ${\bf r}\approx 0$, the duration is
half of the lifetime.  Nevertheless, according to the
results \cite{PRC, freeze-out}, the Landau/Cooper-Frye presentation
of sudden freeze-out could be applied in a generalized form
accounting for momentum dependence of the freeze-out hypersurface
$\sigma_p(x)$; now $\sigma_p(x)$ corresponds to the {\it maximum of
emission function} $S(t_{\sigma}({\bf r},p),{\bf r},p)$ at fixed
momentum ${\bf p}$ in an appropriate region of ${\bf r}$. This
finding allows one to keep in mind the known results
based on the Cooper-Frye formalism, applying them to a surface of
the maximal emission for given $p$. Then the typical features of the
energy dependence can be understood as follows. The inverse of the
spectra slopes, $T_{eff}$, grows with energy, since as one sees from
the emission functions, the duration of expansion increases with
initial energy density and, therefore, the pressure gradient driven fluid elements gets more
transverse collective velocities $v_T$ when reach a decoupling
energy densities. Therefore the blue shift of the spectra becomes
stronger. A rise of the transverse collective flow with energy leads
to some compensation of an increase of $R_{side}$: qualitatively the
homogeneity length at decoupling stage is $R_{side}=
R_{Geom}/\sqrt{1+\langle v_{T}^2\rangle m_{T}/2T}$, (see, e.g.,
\cite{AkkSin}). So, despite an significant increase of the
transverse emission region, $R_{Geom}$, seen in Fig.1, a
magnification of collective flow partially compensates this. It
leads to only a moderate increase of the $R_{side}$ with energy.
Since the temperatures in the regions of the maximal emission
decrease very slowly when initial energy density grows (e.g., the
temperatures for SPS, RHIC and LHC are correspondingly 0.105, 0.103
and 0.95 MeV  for $p_T=0.3$ GeV/c ) the $R_{long}\sim
\tau\sqrt{T/m_T}$ \cite{Averch} grows proportionally to an increase
of the proper time associated with the hypersurface
$\sigma_{p_T}(x)$ of {\it maximal} emission. As we see from Fig. 1
this time grows quite moderate with the collision energy.

A non trivial result concerns the energy behavior of the
$R_{out}/R_{side}$ ratio. It slowly drops when energy grows and
apparently is saturated at fairly high energies at the value close
to unity (Fig.1). To clarify the physical reason of it let us make a
simple half-quantitative analysis. As one can see in Fig. 1, the
hypersurface of the maximal emission can be approximated as
consisting of two parts: the "volume" emission ($V$) at $\tau
\approx const$ and "surface" emission ($S$). A similar picture
within the Cooper-Frye prescription, which generalizes the
blast-wave model \cite{blast-wave} by means of including of the surface emission has
been considered in Ref. \cite{Marina}. If the hypersurface of
maximal emission $\tilde{\tau}(r)$ is double-valued function, as in
our case, then at some transverse momentum $p_T$ the transverse
spectra and HBT radii will be formed mostly by the two contributions
from the different regions with the homogeneity lengths
$\lambda_{i,V}=\sqrt{<(\Delta r_i)^2>}$ ($i$ = side, out) at the
$V$-hypersurface and with the homogeneity lengths $\lambda_{i,S}$ at
the S-hypersurface. Similar to Ref.\cite{AkkSin},  one can apply at
$m_T/T\gg1$ the saddle point method when calculate the single and
two particle spectra using the boost-invariant measures
$\mu_V=d\sigma^V_{\mu}p^{\mu}= \widetilde{\tau}(r)r dr d\phi d\eta
(m_T\cosh(\eta-y)-p_T\frac{d\widetilde{\tau}(r)}{dr}\cos(\phi -
\alpha))$ and $\mu_S=d\sigma^S_{\mu}p^{\mu}= \widetilde{r}(\tau)
\tau d\tau d\phi d\eta
(-m_T\cosh(\eta-y)\frac{d\widetilde{r}(\tau)}{d\tau}+p_T\cos(\phi -
\alpha))$ for $V$- and $S$- parts of freeze-out hypersurface
correspondingly (here $\eta$ and $y$ are space-time and particle
pair rapidities, the similar correspondence is for angles $\phi$ and
$\alpha$, also note that
$\frac{p_T}{m_T}>\frac{d\widetilde{r}(\tau)}{d\tau}$ \cite{PRC,
freeze-out}). Then one can write, ignoring for simplicity the
interference (cross-terms) between the surface and volume
contributions,
\begin{eqnarray}
R_{side}^2 = c_V^2\lambda_{side,V}^2+c_S^2\lambda_{side,S}^2 \label{3} \\
R_{out}^2 = c_V^2\lambda_{out,V}^2+c_S^2\lambda_{out,S}^2(1-
\frac{d\tilde{r}}{d\tau})^2, \label{4}
\end{eqnarray}
where the coefficients $c_V^2+c_S^2\leq1$ and we take into account
that at $p^0/T\gg1$ for pions $\beta_{out}=p_{out}/p^0 \approx 1$.
All homogeneity lengths depends on mean transverse momentum of the
pion pairs $p_T$. The slope $\frac{d\tilde{r}}{d\tau}$ in the region
of homogeneity expresses the strength of $r-\tau$ correlations
between the space and time points of particle emission at the
$S$-hypersurface $\tilde{r}(\tau)$. The picture of emission in Fig.
1 shows that when the energy grows the correlations between the time
and radial points of the emission becomes positive,
$\frac{d\tilde{r}}{d\tau}> 0$, and they increase with energy
density. The positivity is caused by the initial radial flows
\cite{sin1} $u^r(\tau_0)$, which are developed at the pre-thermal
stage, and the strengthening of the $r-\tau$ correlations happens
because the non-central $i$th fluid elements, which produce after
their expansion the surface emission, need more time
$\tau_i(\epsilon_0)$ to reach the decoupling density if they
initially have higher energy density $\epsilon_0$. (Let us
characterized this effect by the parameter
$\kappa=\frac{d\tau_i(\epsilon_0)}{d\epsilon_{0}} > 0 $). Then the
fluid elements before their decays run up to larger radial
freeze-out position $r_i$: if $a$ is the average Lorentz-invariant
acceleration of those fluid elements during the system expansion,
then roughly for $i$th fluid elements which decays at time $ \tau_i$
we have at $a\tau_i \gg 1$: $r_i(\tau_i)\approx
r_i(\tau_0)+\tau_i+(u_i^r(\tau_0)-1)/a$. Then the level of $r-\tau$
correlations within the homogeneous freeze-out "surface" region,
which is formed by the expanding matter that initially at $\tau_0$
occupies the region between the transversal radii $r_1(\tau_0)$ and
$r_2(\tau_0)>r_1(\tau_0)$, is
\begin{equation}
 \frac{d\tilde{r}}{d\tau} \approx \frac{r_1(\tau_1)-r_2(\tau_2)}{\tau_1-\tau_2}
 \approx 1-\frac{R}{\epsilon_0\kappa}\label{5}
\end{equation}
and, therefore, the strength of $r-\tau$ correlations grows with
energy: $\frac{d\tilde{\tau}}{dr}\rightarrow 1$. Note that here we
account for $\tau_2 - \tau_1 \approx \kappa(\epsilon_0(r_2(\tau_0))
- \epsilon_0(r_1(\tau_0)))$ and that
$\frac{d\epsilon_0(r)}{dr}\approx -\frac{\epsilon_0}{R}$ where
$\epsilon_0\equiv\epsilon_0(r=0)$ and $R$ is radius of nuclear. As a
result the second S-term in Eq. (\ref{4}) tends to zero at large
$\epsilon_0$ , reducing, therefore, the $R_{out}/R_{side}$ ratio. In
particular, if $\lambda_{side,V}^2 \gg \lambda_{side,S}^2$ then,
accounting for a similarity of the volume emission in our
approximation and in the blast wave model, where as known
$\lambda_{side,V} \approx \lambda_{out,V}$, one can get:
$\frac{R_{out}}{R_{side}}\approx 1 + const\cdot
\frac{R}{\epsilon_0\kappa}\rightarrow 1$ at $\epsilon_0 \rightarrow
\infty$. It is worthy of note that also measure $\mu_S$ tends to
zero when $\frac{d\tilde{\tau}}{dr}\rightarrow 1$ that again reduces
the surface contribution to $side-$ and $out-$ radii at large $p_T$.

The presented qualitative analysis demonstrates the main
mechanisms responsible for the non-trivial behavior of $R_{out}$ to
$R_{side}$ ratio exposed in  HKM calculations, see Fig.1
(bottom). The very recent first LHC data for Pb+Pb collisions 
presented by the ALICE Collaboration \cite{Alice} conform, in fact, the discussed above
physical picture of space-time evolution responsible for formation of the
HBT radii and  $R_{out}$ to $R_{side}$ ratio, see Fig.1. The transverse femtoscopy scales, predicted for the charged multiplicity
$dN_{ch}/d\eta$=1500 in HKM at the initial energy density $\epsilon_0=40$ GeV/fm$^3$,  are quite close to the experimental data associated with $dN_{ch}/d\eta\approx 1600$ at the collision energy $\sqrt{s}=2.76$ TeV. As for the longitudinal HBT radius, $R_{long}$, it is underestimated  in HKM by around 20\%. As the result, HKM gives smaller interferometry volume than is observed at LHC. The reason could be that HKM describes a gradual decay of the system which evolves hydrodynamically until fairly large times. It is known \cite{AkkSin2} that at the isentropic and chemically frozen hydrodynamic evolution the interferometry volume increases quite moderate with initial energy density growth in collisions of the same/similar nucleus. The RHIC results  support such a theoretical view (see solid line in Fig.2), while the ALICE Collaboration observes a significant increase of the interferometry volume at LHC. One should change, thus, the global fit of $V_{int}(dN/d\eta)$ for A+A collisions for steeper slope (upper dash line). However, no one linear fit cannot be extrapolated to $V_{int}(dN/d\eta)$-dependence discovered by the ALICE Collaboration in p+p collisions \cite{Alice2} (bottom dashed line in Fig.2). Could one call these two observed peculiarities as the "`LHC HBT puzzle"'? On our opinion, at least qualitavely, it is not puzzling. An essential growth  of the interferometry volume in Pb+Pb collisons at the first LHC energy can be conditioned by an increase of the duration of the last very non-equilibrium stage of the matter evolution which cannot  be considered on the hydrodynamic basis and one should use hadronic cascade models like UrQMD. At such late stage the results obtained in \cite{AkkSin2} for isentropic and chemically frousen evolution are violated. As for the different linear $V_{int}(dN/d\eta)$ dependence in A+A and p+p collisions, the interferometry volume depends not only on multiplicity but also on intial size of colliding systems \cite{AkkSin2}. Therefore, qualitavely, we see no puzzle in the newest HBT results obtained at LHC in Pb+Pb and p+p collisions, but the final concusion can be done only after detailed quantitative analysis.

{\it Summary}---We conclude that energy behavior of the pion
interferometry scales can be understood at the same hydrokinetic basis as for the SPS and RHIC energies supplemented by hadronic cascade model at the latest stage of the evolution.
In this approach the EoS  accounts for a crossover
transition between quark-gluon and hadron matters at high collision
energies and non-equilibrated expansion of the
hadron-resonance gas at the later stage.\\
\begin{figure*}[htb]
\centering
 \includegraphics[scale=0.7]{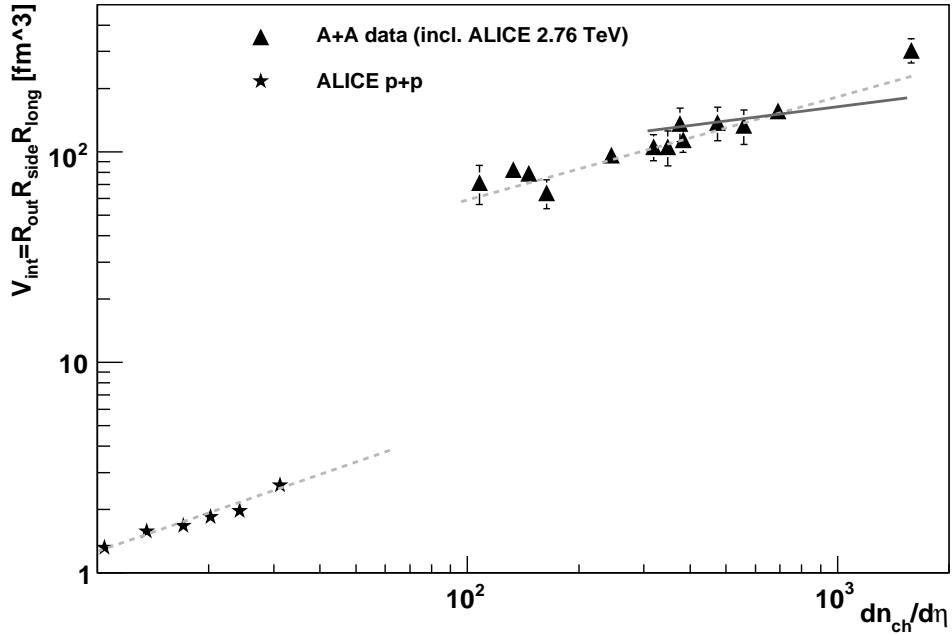}%
 
 \vspace{-0.1in}
\caption{Illustration of  multiplicity dependence of the pion inteferometry volume on charged particle multiplicity for central heavy ion collisions at  AGS, SPS, RHIC and LHC energies and comparison with the results in p+p collisions (bottom, left). All the HBT radii are taken at the pion transverse momentum $p_T=$ 0.3 GeV. For A+A collision the data are taken from Fig. 4 of Ref. \cite{Alice} (see all details there), for p+p collisions the points for  $p_T=$ 0.3 GeV are interpolated from the results of Ref. \cite{Alice2}. The solid line corresponds to linear fit of $V_{int}(dN_{ch}/d\eta)$-dependence only for the top SPS and the RHIC energies; upper dashed line is fit for all the A+A energies including the newest LHC point at $\sqrt{s}=2.76$ TeV; the bottom dashed line is the linear fit for ALICE LHC results for p+p collisions with energies 0.9 and 2.76 TeV.} 
\end{figure*}

The HKM alows one to treat correctly the process of particle
emission from expanding fireball, that is not sudden and lasts about
system's lifetime. Also it takes into account  the prethermal
formation of transverse flows.  Then the
main mechanisms that lead to the paradoxical behavior of the
interferometry scales find a natural explanation. In particular, a
slow decrease and apparent saturation of $R_{out}/R_{side}$ ratio
around unity at high energy happens due to a strengthening  of
positive correlations between space and time positions of pions
emitted at the radial periphery of the system. Such an effect is a
consequence of the two factors accompanying an increase of collision
energy: a developing of the pre-thermal collective transverse flows
and an increase of initial energy density in the fireball. The prediction of the HKM for LHC energies are quite close to the first experimental data in Pb+Pb collisions at LHC. 
\section*{Acknowledgments}
 Yu.S. gives thanks to P. Braun-Munzinger for support this study within EMMI/GSI organizations as well as
 for fruitful and very stimulating discussions. The researches were carried 
  out in part within the scope of the EUREA: European Ultra Relativistic Energies Agreement (European Research Group
  GDRE: Heavy ions at ultrarelativistic energies) and is supported by the State Fund for
Fundamental Researches of Ukraine (Agreement of 2011)  and National Academy of Sciences of Ukraine (Agreement of 2011).


\begin{thebibliography}{99}

\bibitem{LHCpred} \textit{Armesto N. (ed. ), et al.} // J. Phys. 2008. V.35. P.054001.
\bibitem{Bertsch} \textit{Bertsch G.}// Phys. Rev. C. 1989. V.40. P.1830; \textit{Rischke D.H., Gyulassy M.}//Nucl. Phys. A 1996. V. 608. P.479.
\bibitem{Sin}\textit{Sinyukov Yu.M.}// Nucl.Phys. A. 1994. V. 566. P.589c;
\textit{Sinyukov Yu.M.}//Hot Hadronic Matter: Theory and Experiment,
eds. J. Letessier, H.H. Gutbrod and J. Rafelski (Plenum, New York)
1995,P. 309.
\bibitem{HBTpuzzle} \textit{Heinz U.}// Nucl. Phys. A 2003. V.721. P.30; \textit{Pratt S.}// Nucl. Phys. A 2003. V.715. P.389c; \textit{ Soff S., Bass S., Hardtke D., Panitkin S.}// Nucl. Phys. A. 2003. V.715. P. 801c.
\bibitem{sin1} \textit{Sinyukov Yu.M.} // Acta Phys. Polon. B. 2006. V.37. P.4333;//
\textit{Gyulassy M. et al.} // Braz. J. Phys. 2007. V.37. P.1031.
\bibitem{PRC} \textit{Akkelin S.V., Hama Y., Karpenko Yu.A., Sinyukov Yu.M.}// Phys. Rev. C. 2008. V.78. P.034906.
\bibitem{JPG} \textit{Sinyukov Yu.M., Karpenko Iu.A.,Nazarenko A.V.}//
 J. Phys. G: Nucl. Part. Phys. 2008. V. 35. P. 104071.
\bibitem{Cracow}\textit {Broniowski W.,  Florkowski W.,Chojnacki M.,, Kisiel A.}// Phys. Rev. C 2009.V.80. P 034902.
\bibitem{sin2}\textit {Sinyukov Yu.M., Nazarenko A.V., Karpenko Iu.A.}// Act. Phys. Pol. B. 2009.  V.40. P. 1109.
\bibitem{Pratt}\textit {Pratt S.}// Nucl.Phys.A 2009. V.830. P. 51c-57c. 
\bibitem{sin3}\textit{Karpenko Iu.A., Sinyukov Yu.M.}//Phys.Rev. C. 2010. V. 81. P. 054903.


\bibitem{Alice}\textit{Aamodt K., et al (ALICE Collaboration)}//Phys. Lett. B. 2011. V. 696. P.328.
\bibitem{sin4}\textit{Karpenko Iu.A., Sinyukov Yu.M.}//Phys. Lett. B. 2010. V. 688. P. 50.

\bibitem{PRL} \textit{ Sinyukov Yu.M., Akkelin S.V.,  Hama Y.}// Phys. Rev.
Lett. 2002. V. 89. P. 052301.
\bibitem{Kolb} \textit{Kolb P. F., Sollfrank J., Heinz U.}// Phys. Lett. B. 1999. V. 459. P.
667; and Phys. Rev. C. 2000. V. 62. P.054909.


\bibitem{Teaney} \textit{Teaney D.}// Phys. Rev. C. 2003. V. 68. P. 034913.
\bibitem{Laine} \textit{Laine M., Schr\"{o}der Y.} Phys. Rev. D. 2006. V.73. P. 085009.
\bibitem{Toneev}\textit{Karpenko~Iu.A., Khvorostukhin~A.S.,  Toneev~V.D., Sinyukov~Yu.M.}//arXiv 2010 [nucl-th]  N.1012.2312.

\bibitem{PBM1}\textit {Becattini F., Manninen J.}// J. Phys. G: Nucl. Part. Phys. 2008. V.35. P.
104013; \textit {Andronic A., Braun-Munzinger P. , Stachel J. }//
 arXiv 2008:0812.1186; arXiv 2009:0901.2909.
\bibitem{UrQMD} \textit {Bleicher M. et al.}// J. Phys. G: Nucl. Part. Phys. 1999. V. 25. P.
1859.
\bibitem{freeze-out} \textit{Sinyukov Yu.M., Akkelin S.V., Karpenko Iu.A.}// Acta Phys.Polon.B. 2009. V.40.
P.1025.
\bibitem{AkkSin} \textit{Akkelin S.V., Sinyukov Yu.M.}// Phys. Lett. B 1995. V. 356. P.
525; \textit{Akkelin S.V., Sinyukov Yu.M.}// Z. Phys. C. 1996. V.72. P. 501.
\bibitem{Averch} \textit{Makhlin A. N., 
Sinyukov Yu. M.}// Z. Phys. C. 1988. V.39 P.69.
\bibitem{blast-wave} \textit {Schnedermann E., Sollfank J., Heinz U.}// Phys. Rev. C. 1993. V. 48. P. 2462.
\bibitem{Marina} \textit {Borysova M.S., Sinyukov Yu.M. , Akkelin S.V. , Erazmus B. , Karpenko Iu.A.}//
Phys.Rev. C. 2006. V.73. P. 024903.
\bibitem {AkkSin2}\textit{Akkelin S.V., Sinyukov Yu.M.}// Phys.Rev. C. 2004. V. 70.  P. 064901; Phys.Rev. C. 2006. V. 73.  P. 034908 
\bibitem{ceres} \textit{Anto\'nczyk Dariusz}// Acta Phys. Polon. B. 2009. V. 40. P. 1137.
\bibitem{na49-spectra} \textit{ Afanasiev S. V. et al, NA49 Collaboration}// Phys. Rev. C. 2002. V. 66. P. 054902.
\bibitem{na49-hbt} \textit{Alt C.  et al, NA49 Collaboration}// Phys. Rev. C. 2008. V.77. P. 064908.

\bibitem{star-spectra} \textit{Adams J. et al. (STAR Collaboration)} // Phys. Rev. Lett. 2004. V. 92. P. 112301.
\bibitem{star-hbt} \textit{Adams J. et al. (STAR Collaboration)} // Phys. Rev. C. 2004. V.71. P. 044906.
\bibitem{phenix-spectra} \textit{Adler S.S. et al. (PHENIX Collaboration)} // Phys. Rev. C. 2004. V. 69. P. 034909.
\bibitem{phenix-hbt} \textit{Adler S.S. et al. (PHENIX Collaboration)}// Phys. Rev. Lett. 2004. V. 93. P. 152302.
\bibitem{Alice2}\textit{The ALICE Collaboration}// arXiv [hep-ex] 2011:1101.3665
\end{thebibliography}
\end{document}